\documentstyle[12pt,emlines]{article}
\begin{document}
\author{L.Z. Kon}
\title{Some kinetic properties of the two-band superconductors}\maketitle
Institute of Applied Physics, Chishinau, Moldova

\begin{abstract}
In the comment of M.E. Palistrant (arXiv:\thinspace
cond-mat\thinspace/0305496\thinspace ) the main results of the theory of
superconductors with overlapping energy bands are formulated. The comment
provides a list of references to the classical papers of the author of this
theory, Prof. V.A.Moskalenko, and coworkers. This theory has been generalized by
Prof. M.E.Palistrant and coworkers to the case of a reduced density of charge
carriers. Respective references are also included in the comment.

In our opinion, it would be useful to provide in addition to the above comment
the references covering our research of the kinetic properties of
superconductors with overlapping energy bands. In particular, we mention some
important results obtained many years ago: the Ginsburg-Landau (GL)equations for
the two-band superconductors doped with impurities and the influence of
impurities on the energy gap, as well as those concerning dynamical properties
of the two-band superconductors.
\end{abstract}

The discovery and experimental investigation of superconducting properties of
MgB$_{2}$ have attracted a special interest to the model with overlapping energy
bands. In some cases, this model is extended to take into account different
anisotropies of the order parameters, as well as the strong electron-lattice
coupling. The two-band model is used to interpret experimental data on
tunneling, specific heat, electronic Raman scattering, thermal conductivity,
penetration depth of the magnetic field and other properties of MgB$_{2}$.

Prof. V.A. Moskalenko and his coworkers from the Institute of Applied Physics of
the Academy of Sciences of Moldova have studied most of these properties on the
basis of the above model. The equilibrium problem is described by the two-band
Hamiltonian \cite{1}. This Hamiltonian has been extended to consider the
scattering of electrons by non-magnetic impurities \cite{2}. The results for
pure and doped systems are valid for arbitrary values of the two-band
parameters.

 We present here only some qualitatively new kinetic properties, which have been
obtained for the model with overlapping energy bands.

1. The GL system of equations for the two-band model has been formulated to
cover the whole range of parameters from pristine to the dirty limit. On this
basis, the magnetic penetration depth of a superconductor, the jump of the
specific heat per unit cell at the critical temperature, and other properties
have been investigated in \cite{3},\cite{4},\cite{5}.

For a high concentration of non-magnetic impurities, the system of GL equations
for the two-band model is similar to the GL system of the one-band model.
However, the critical temperature T$_{c}$ and the dimensionless GL-parameter
$\kappa $ in the equations are determined by the two-band model. In this case
the expression for the relative jump of the specific heat at T$_{c}$ coincides
with the corresponding expression for the pristine one-band model, with density
of states being the sum of densities of the two bands and the dependence of
T$_{c}\,$on impurities specified by the two band model. The effect of magnetic
impurities is to decrease the relative specific heat jump and to increase
considerably both the parameter $\kappa $and the magnetic field penetration
depth in the superconducting phase.

2. The influence of impurities (non-magnetic and paramagnetic) on the
thermodynamic properties of the two- band superconductors at zero, close to zero
and at the critical temperatures has been considered in
\cite{6},\cite{7},\cite{8}. It has been found that due to interband scattering
of electrons on impurities, the superconducting state of the two-band model is
described by a single energy gap. Thus, when one of the densities of states be
comes zero, the other density vanishes too. In particular, the energy gap in the
"dirty" limit for a non-magnetic impurity decouples in a product of averages of
the order parameters of individual bands with their densities serving as weight
factors.

3. The non-equilibrium process of charge imbalance in a two-band superconductor
has been investigated by employing of the Keldysh Green functions technique. The
kinetic equation, the penetration depth of the longitudinal electric field and
the distribution of this field in the superconductor are given in \cite{9}.

A new mechanism of relaxation of the charge imbalance in non-equilibrium
two-band superconductors has been revealed. This mechanism is due to interband
electron-impurity scattering and leads to a decrease of the penetration depth of
the longitudinal electric field into the superconductor.

4. A model of a superconductor with two dielectric gaps and two superconducting
order parameters corresponding to the two parts of the Fermi surface has been
formulated. The phase diagram obtained for this model contains an area of
coexistence of structural, antiferromagnetic and superconducting phase
transitions versus the non-magnetic impurity concentration that agrees
qualitatively with the experimental data on high-temperature supercondactors
\cite{10}.

5. In \cite{11} the electronic Raman scattering in superconductors, taking into
account the collective oscillations, Coulomb screening, and scattering of
electrons by non-magnetic impurities has been studied in the framework of the
two-band model.

Two contributions to the intensity of the scattered light have been singledout:
an additive contribution from each of the two bands, and a term caused by the
interband transitions of Cooper pairs which exists for an arbitrary light
polarization. Experimentally, this means that the lowest gap should beactive for
any light polarization.

6. The propagation of longitudinal ultrasound in the one- and two-band models of
superconductors at arbitrary temperatures has been investigated by taking into
account the collective oscillations in the presence of non-magnetic impurities
for an arbitrary mean free path. The effect of superconductivity and impurity on
the relative shift of the sound velocity turned out to depend strongly on the
choice of the model \cite{12},\cite{13}.

In particular, we have predicted a more efficient suppression by impurities of
fluctuations of the superconducting gaps for the two-band model,  than for the
one-band model. The two-band model has also allowed us to explain such
aspectacular feature of high-T$_{c}$ superconductors as the increase of sound
velocity for all the temperature interval below T$_{c}$.

7. For a two-band superconductor the amplitude of multiple electron scattering
by non-magnetic impurities does not have electron-hole symmetry with respect to
the Fermi surface, and this may be the cause of an increase in the
thermoelectric effect in superconductors. As a result, the temperature
dependence of the additional contribution to thermoelectric coefficient reaches
a maximum in the region of temperatures $T<T_{c}$ $\,$

8. The collective modes related to phase fluctuations near T$_{c}$ have been
investigated by assuming the existence of a two-component neutral superfluid.
The equation for collective modes describes interference of the two processes:
small fluctuations of relative density of the
condensate of electrons (Leggett-type)and small fluctuations
of the charge imbalance of the electron-hole branches. This equation is
analogous to the well-known type of equation in solid state physics describing,
e.g., the collective modes of polaritons \cite{15}.

The amplitude collective modes of the two-band model have also been studied. We
mention that these modes in the case of non-identical traditional two-band
superconductors do not occur$\,$\cite{16}

\end{document}